\documentclass{article}
\usepackage{emulateapj}

\def\etal{{\it et al. }}
\def\kms{\rm ~km~s^{-1}}

\def\simlt{\mathrel{\spose{\lower 3pt\hbox{$\mathchar"218$}}
     \raise 2.0pt\hbox{$\mathchar"13C$}}}
\def\simgt{\mathrel{\spose{\lower 3pt\hbox{$\mathchar"218$}}
'     \raise 2.0pt\hbox{$\mathchar"13E$}}}
\def\gsim{ \lower .75ex \hbox{$\sim$} \llap{\raise .27ex \hbox{$>$}} }
\def\lsim{ \lower .75ex \hbox{$\sim$} \llap{\raise .27ex \hbox{$<$}} }

\accepted{}

\long\def\***#1{{\scshape ***#1***}}
\def\arraystretch{1.25}

\makeatletter

\newenvironment{inlinefigure}{%
\def\@captype{figure}%
\noindent\begin{minipage}{0.999\linewidth}\begin{center}}
{\end{center}\end{minipage}\smallskip}
\makeatother

\begin{document}
\lefthead{MOORE ET AL.}
\righthead{DARK MATTER SUBSTRUCTURE}
\submitted{Astrophysical Journal Letters, in press.}

\title{Dark matter substructure within galactic halos} 
  \author{B.~Moore\altaffilmark{1},
  S.~Ghigna\altaffilmark{1}, F.~Governato\altaffilmark{1,4},
  G.~Lake\altaffilmark{2}, T.~Quinn\altaffilmark{2},
  J.~Stadel\altaffilmark{2}, P.~Tozzi\altaffilmark{3}} 

\altaffiltext{1}{Physics Department, University of Durham, Durham
City, UK, DH1 3LE}

\altaffiltext{2}{Astronomy Department, University of Washington,
Seattle, USA, WA 98195}

\altaffiltext{3}{
Osservatorio Astronomico di Roma, Italy
\& John Hopkins University Baltimore, USA} 

\altaffiltext{4}{Osservatorio Astronomico di Brera, Merate, Italy}

\begin{abstract}
We use numerical simulations to examine the substructure within
galactic and cluster mass halos that form within a hierarchical
universe. Clusters are easily reproduced with a steep mass spectrum of
thousands of substructure clumps that closely matches observations.
However, the survival of dark matter substructure also occurs on
galactic scales, leading to the remarkable result that galaxy halos
appear as scaled versions of galaxy clusters.  The model predicts that
the virialised extent of the Milky Way's halo should contain about 500
satellites with circular velocities larger than Draco and Ursa-Minor
{\it i.e.} bound masses $\gsim 10^8M_\odot$ and tidally limited sizes
$\gsim$ kpc.  The substructure clumps are on orbits that take a large
fraction of them through the stellar disk leading to significant
resonant and impulsive heating.  Their abundance and singular density
profiles has important implications for the existence of old thin
disks, cold stellar streams, gravitational lensing and indirect/direct
detection experiments.
\end{abstract}


\section{Introduction}

The growth of structure in the universe by hierarchical accretion and
merging of dark matter halos is an attractive and well motivated
cosmological model (White \& Rees 1978, Davis \etal 1985).  The
gravitational clustering process is governed by the dark matter
component and the baryons only play a minor role.  The idea that
galaxies are defined as those objects where gas can quickly cool
predates the current hierarchical model (Hoyle 1953), and has been
invoked to set the scale for survival versus disruption (Rees \&
Ostriker 1977, White \& Rees 1978).

It has proved difficult to compare the predictions of this model with
non-linear structures, such as the internal properties of galaxy
clusters.  Numerical simulations had ubiquitously failed to find
surviving substructure or ``halos orbiting within halos'' (e.g. Katz
\& White 1993, Summers \etal 1995, Frenk \etal 1996).  It
was generally thought that the so called ``over-merging'' problem
could be overcome by the inclusion of baryonic component to increase
the potential depth of galactic halos.

Analytic work suggested that over-merging was due entirely to poor
spatial and mass resolution (Moore, Katz \& Lake 1996).  This has been
verified by higher resolution simulations of clusters in which
galactic halos survive without any inclusion of gas dynamics (Moore
\etal 1998, Ghigna \etal 1998, Klypin \etal 1998). When a galaxy and
its dark matter halo enter a larger structure, the outer regions are
stripped away by the global tides and mutual interactions.  The
central region survives intact so that a galaxy may continue to be
observed as a distinct structure within a cluster, with its own
truncated dark matter halo (Natarajan \etal 1998).

In a hierarchical universe, galaxies form by a similar merging and
accretion process as clusters (Klypin \etal 1999).  Over-merging on
galactic scales is a necessary requirement otherwise previous
generations of the hierarchy would preclude the formation of disks.
Observations ssuggest that over-merging has been nearly
complete on galactic scales. The Milky Way contains just 11 satellites
within its virial radius with $\sigma_{satellite}/\sigma_{halo}\gsim
0.07$, i.e. that is equivalent to $\sigma_{satellite}=10\kms$
(c.f. Mateo 1998 and references within).  The same velocity dispersion
ratio in a cluster corresponds to counting galaxies more massive than
the Large Magellanic Clouds $\sigma_{LMC} \sim 50 \kms$; there are
500-1000 such systems in a rich cluster (Binggeli \etal
1985, Driver \etal 1999).  The same discrepancy exists
at higher masses.  The Coma cluster contains $\gsim 30$ galaxies
brighter than the characteristic break in the luminosity function,
$L_* \equiv \sigma>200\kms$ (Lucey \etal 1991).  Scaling this limit to
a galaxy halo we find just 2 satellites in the Milky Way or 3 near
Andromeda.

Why should substructure be destroyed in galactic halos but not in
clusters?  Analytic calculations suggested that galaxies should
contain more satellites than observed (Kauffmann \etal 
1993).  The shape of the power spectrum varies over these scales such
that galaxies form several billion years before clusters and the mass
function of their progenitor clumps may differ.  Furthermore, as the
power spectrum asymptotically approaches a slope of -3, clumps of all
masses will be collapsing simultaneously and the timescale between
collapse and subsequent merging becomes shorter.  These effects may
conspire to preferentially smooth out the mass distribution within
galactic halos.  In this {\it letter} we use numerical simulations to
study the formation of galactic halos with sufficient force and mass
resolution that can resolve satellites as small as Draco.  
This allows us to make a comparative study with
observations and simulations of larger mass halos.

\vfil\eject

\section{Substructure within galaxies and clusters}

We simulate the hierarchical formation of dark matter halos in the
correct cosmological context using a high resolution parallel {\sc
treecode pkdgrav}.  An object is chosen from a simulation of an
appropriate cosmological volume.  The small scale waves of the power
spectrum are realised within the volume that collapses to this object
with progressively lower resolution at increasing distances from the
object. The simulation is then re-run to the present epoch with the
higher mass and force resolution.  We have applied this technique to
several halos identified from a $10^6$ Mpc$^3$ volume, including a
cluster similar to the nearby Virgo cluster (Ghigna \etal 1998) and a
galaxy with a circular velocity and isolation similar to the Milky
Way.

\begin{inlinefigure}
\bigskip
\centerline{\includegraphics[width=1.0\linewidth]{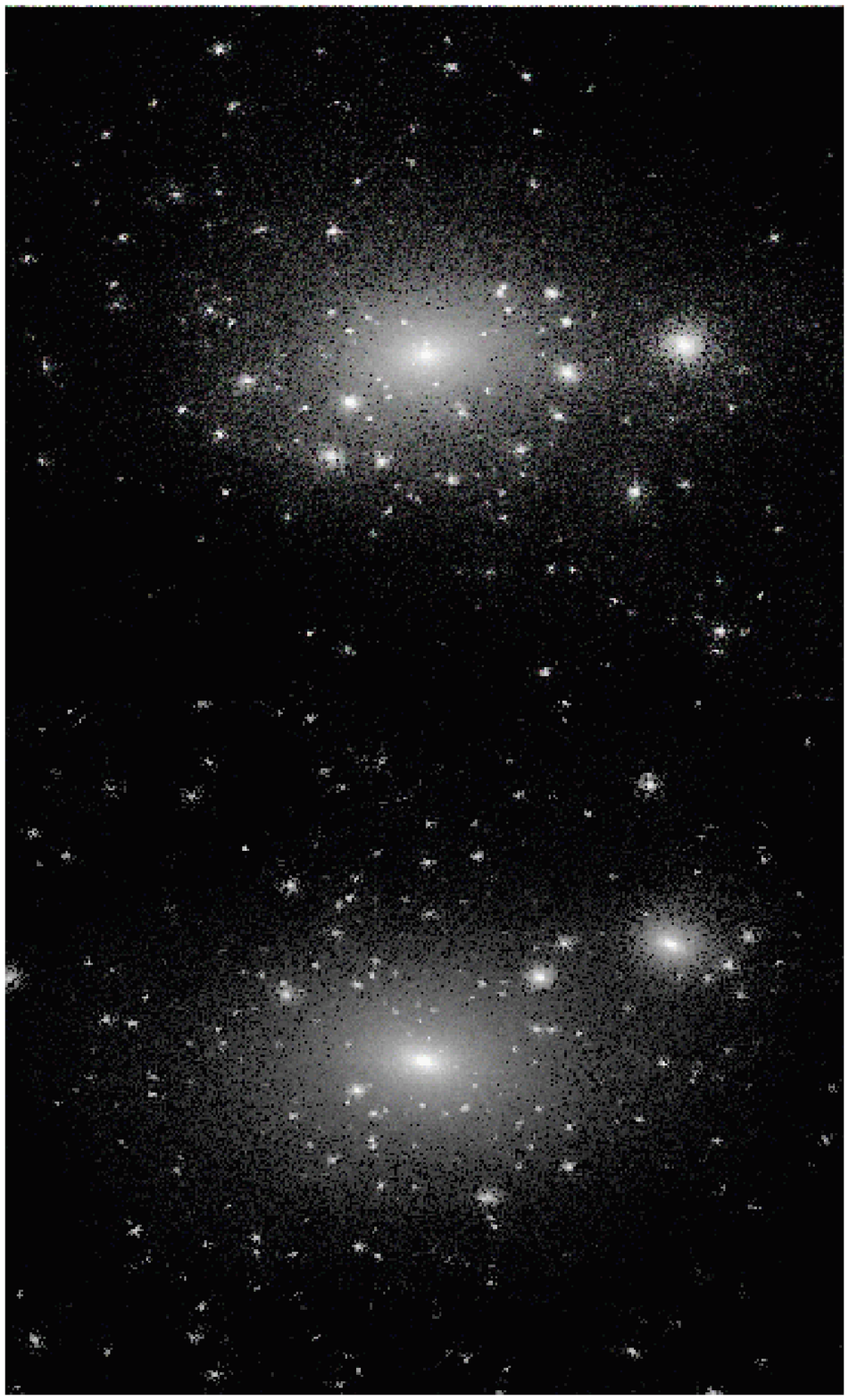}}
\bigskip
\caption{ The density of dark matter within a cluster halo of mass
$5\times 10^{14}M_\odot$ (upper) and a galaxy halo of mass $2\times
10^{12}M_\odot$ (lower).  The edge of the box is the virial radius,
300kpc for the galaxy and 2000 kpc for the cluster (peak circular
velocities of $200 \kms$ and $1100 \kms$ respectively).
}
\label{fig:1159:opt}
\end{inlinefigure}

The cosmology that we investigate is a universe dominated with a
critical density of cold dark matter, normalised to reproduce the
local abundance of galaxy clusters.  The important numerical
parameters to remember are that each halo contains more than one
million particles within the final virial radius $r_{vir}$, and we use
a force resolution $\sim 0.1\% r_{vir}$. Further details of
computational techniques and simulation parameters can be found in
Ghigna \etal (1998) and Moore \etal (1999). Here we focus our
attention directly on a comparison with observations.

Figure 1 shows the mass distribution at a redshift $z=0$ within the
virial radii of our simulated cluster and galaxy. It is virtually
impossible to distinguish the two dark matter halos, even though the
cluster halo is nearly a thousand times more massive and forms 5 Gyrs
later than the galaxy halo. Both objects contain many dark matter
substructure halos.
We apply a group finding algorithm to extract the sub-clumps from the
simulation data and use the bound particles to directly measure their
kinematical properties; mass, circular velocity, radii, orbital
parameters ({\it c.f.} Ghigna \etal 1998).  Although our simulations
do not include a baryonic tracer component, we can compare the
properties of these systems with observations using the Tully-Fisher
relation (Tully \& Fisher 1977).  This provides a simple benchmark for
future studies that incorporate additional physics such as cooling gas
and star-formation.

\begin{inlinefigure}
\centerline{\includegraphics[width=1.0\linewidth]{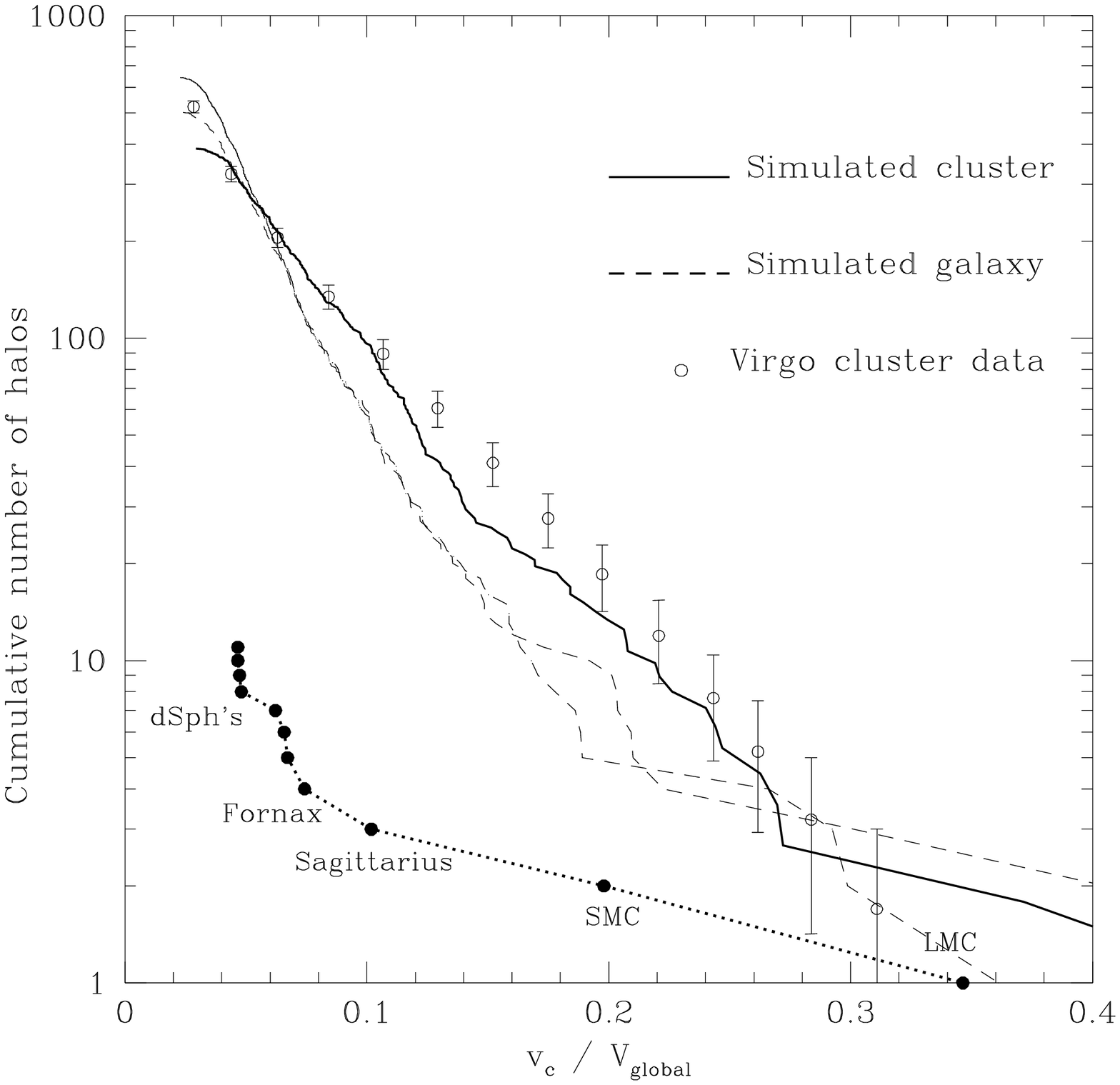}}
\caption{
The abundance of cosmic substructure within our Milky Way Galaxy, the
Virgo cluster and our models of comparable masses.  We plot the
cumulative numbers of halos as a function of their circular velocity
($v_c=\sqrt{(Gm_b/r_b)}$, where $m_b$ is the bound mass within the
bound radius $r_b$ of the substructure, normalised to the circular
velocity, $V_{global}$ of the parent halo that they inhabit. The
dotted curve shows the distribution of the satellites within the Milky
Way's halo (Mateo 1998) and the open circles with Poisson errors is
data for the Virgo galaxy cluster (Binggeli \etal 1985).
We compare these data with our simulated galactic mass halo (dashed curves) and
cluster halo (solid curve).  The second dashed curve shows data for the galaxy
at an earlier epoch, 4 billion years ago \-- dynamical evolution has not
significantly altered the properties of the substructure over this timescale.
}
\label{fig:1159:opt}
\end{inlinefigure}

Figure 2 shows the observed mass (circular velocity) function of
substructure within the Virgo cluster of galaxies compared with our
simulation results.  The circular velocities of substructure halos are
measured directly from the simulation, whilst for the Virgo cluster we
invert the Binggeli \etal luminosity function data using the
Tully-Fisher relation.  There are no free parameters to this fit.  The
overall normalization of the simulation was fixed by large scale
clustering properties and we then picked a cluster from a low
resolution run that had a dispersion similar to Virgo.  We count as a
remarkable success that this model reproduces both the shape and
amplitude of the galaxy mass function within a cluster.

Also plotted in Figure 2 is the cumulative distribution of the 11
observed satellites that lie within 300 kpc of the Milky Way.  Where
necessary we have converted 1d velocity dispersions to circular
velocities assuming isotropic velocity distributions. The model
over-predicts the total number of satellites larger than the dSph's by
about a factor of 50.

The distribution of circular velocities for the model galaxy and
cluster can be fitted with a power law $n(v/V_{vir}) \propto
(v/V_{vir})^{-4}$, similar to that found by Klypin \etal (1999) for
satellites in the proximity to galactic halos.  The mass function
within these systems can be approximated by a power law with
$n(m/M_{vir}) \propto (m/M_{vir})^{-2}$.  The
tidally limited substructure halos have profiles close to isothermal
spheres with core radii equal to our resolution length - increasing
the resolution only makes the halos denser and more robust to
disruption (Moore \etal 1998).

\section{Discussion}

Either the hierarchical model is fundamentally wrong, or the
substructure lumps are present in the galactic halo and contain too
few baryons to be observed.  The deficiency of satellites in galactic
halos is similar to a deficiency of dwarf galaxies in the field
(e.g. Kauffmann \etal 1993).
One possibility is that some of the missing
satellites may be linked to the high velocity clouds (Blitz \etal
1999). Numerous studies have invoked feedback from star formation or an
ionizing background to darken dwarfs by expelling gas and inhibiting
star formation in low mass halos (Dekel \& Silk 1986; Quinn, Katz \&
Efstathiou 1996).  The case for feedback has always been weak.
Galaxies outside of clusters are primarily rotationally supported
disks, their final structure has clearly been set by their angular
momentum rather than a struggle between gravity and winds.  The
strongest star-bursts seen in nearby dwarf galaxies lift the gas out
of their disks, but the energy input is insufficient to expel the gas
and reshape the galaxy (Martin 1998).

While there might be little consequence to darkening dwarfs in the
field, spiral disks will neither form nor survive in the presence of
large amounts of substructure.  The strongly fluctuating potential of
clumpy collapses inhibits disk formation and has been shown to be an
effective formation mechanism for creating elliptical galaxies (Lake
\& Carlberg 1988; Katz \& Gunn 1991; Steinmetz \& Muller 1995).  
Figure 3 shows the proto-galactic mass distribution at a redshift of
10, just a billion years after the big-bang.  The smallest collapsed
halos that we can resolve have a mass of $10^7M_\odot$, not much
larger than globular clusters.  The problem of baryonic trapping by
star formation in lumps arises before the first QSO's could ionize the
intergalactic material (however, see Haiman, Abel \& Rees 1999).  The
second problem is that the lumps do not dissolve by $z=1$ or even by
the present day, as we have shown.  Even if we make the most
optimistic assumptions about the fate of gas, the movements of this
small tracer component will not lead to the destruction of the dark
matter substructure.

\begin{inlinefigure}
\bigskip
\centerline{\includegraphics[width=1.0\linewidth]{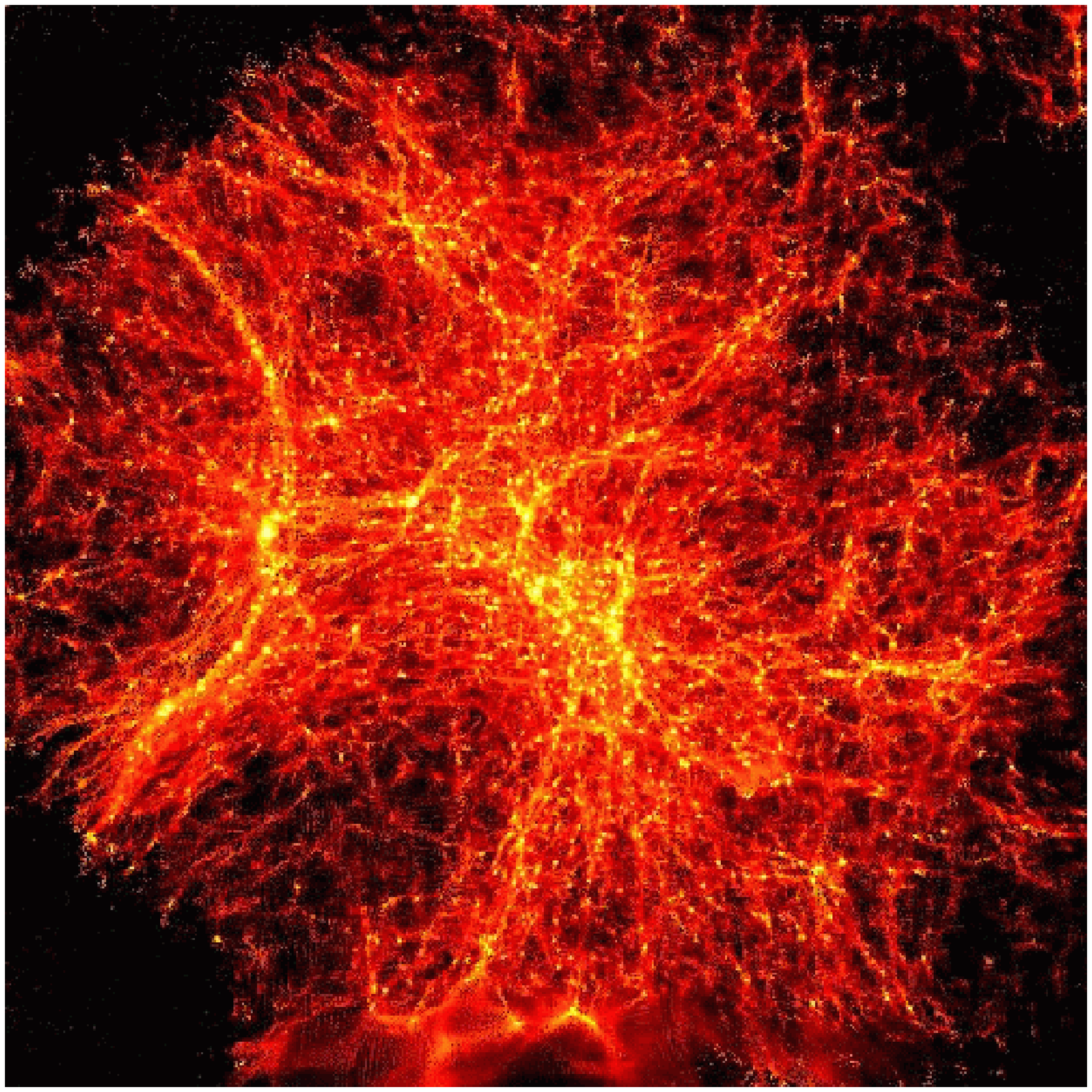}}
\bigskip
\caption{
The distribution of mass at a redshift z=10 in a 6 comoving Mpc region
that forms a galactic halo by the present day.  The colours show the
smoothed local density at the position of each particle plotted in the
range $\delta\rho/\rho=10\--10^6$. The smallest halos that we can
resolve have circular velocities of $10\kms$; virialised halos appear
as bright yellow/white blobs. The cooling time for primordial gas
within these halos is extremely short and leads to the
``over-cooling'' problem: most of the baryons in the universe will be
trapped within low mass halos leaving no gas left to form disks at
late times. }

\label{fig:1159:opt}
\end{inlinefigure}


The most obvious observational 
constraints are the existence of old thin disks (Wielen 1972) and cold
stellar streams (Shang \etal 1998).  Just as gravitational
perturbations from encounters transform disk galaxies to spheroidals
in clusters (Moore \etal 1996), the passage of these lumps will heat
any disk within the halo.  Stellar disks extend to $\sim 10\%$ of the
virial radius of the dark matter halo although HI can be observed to
much larger distances (25\% of $r_{vir}$ for some LSB galaxies).


We find that the orbits of satellites within our simulated halos have
a median apocentric to pericentric distance of 6:1, therefore over the
past 10 billion years disks will suffer many thousands of impulsive
shocks and resonant heating.  The single accretion of a satellite as
large as the Large Magellanic Cloud, has a devastating effect on the
disk of the Milky Way (Toth \& Ostriker 1992; Ibata \etal 1998,
Weinberg 1998).  While recent work has noted that disks embedded
within live halos may precess in response to a single satellite and
avoid strong vertical heating (Huang \& Carlberg 1997; Velazquez \&
White 1998), there are far too many clumps in our simulations for this
mechanism to be effective.

An estimate of the heating can be obtained using the impulse
approximation.  Each dark halo that passes nearby or through the disk,
will increase the stellar velocities across a region comparable to the
size of the perturber by an amount $\delta v \sim G m_b/r_b V$ where
$V$ is the impact velocity. We measure $m_b$, $r_b$ and $V$ for each
clump that orbits through the stellar disk.  Summing the $\delta v$'s
in quadrature over 10 Gyrs, we find that the energy input from
encounters is a significant fraction of the binding energy of the
stellar disk $\sim M_d v_c^2$, where $M_d$ and $v_c$ are the disk mass
and rotation velocity respectively.  The heating is more than
sufficient to explain the age-temperature relation for disk stars
(Wielen 1974), although the validity of the impulse approximation
needs to be examined using numerical simulations. We note that the
existence of old thin disk components, or galaxies such as NGC 4244
that does not have a thick disk (Fry \etal 1999), presents a severe
problem for hierarchical models.

Substructure can be probed by gravitational lensing even if stars are
not visible in the potential wells (e.g. Hogan 1999).  Multiply imaged
quasars are particularly sensitive to the foreground mass
distribution; the quadruple images QSO 1422+231 cannot be modeled with
a single smooth potential (Mao \& Schneider 1998) and requires
distortions of $\approx$ 1\% of the critical surface density within
the Einstein radius.  Dark matter substructure located in projection
near to the primary source would create such distortions. If we
extrapolate our mass function to smaller masses, we expect $\approx
10^5$ clumps with $v_c/V_{200}>0.01$ ($m_b\approx 10^6M_\odot$).  This
may cause many gravitationally lensed quasars to show signs of
substructure within the lensing potentials.

Cold dark matter candidates, such as axions and neutralinos, can be
detected directly in the laboratory. Many proposed and ongoing
experiments will be highly sensitive to the phase space distribution
of particles at our position within the galaxy, yet calculations of
experimental rates still assume that CDM particles passing through
minute detectors have a smooth phase space distribution. We have shown
that CDM halos are far from smooth, furthermore, the particle
velocities in a single resolution element have a discrete component
that results from the coherent streams of particles tidally stripped
from individual dark matter halos. We may also an expect enhanced
gamma-ray flux from neutralino annihilation within substructure cores
(Lake 1990, Bergstrom \etal 1998).

\section{Summary}

In a hierarchical universe, galaxies are scaled versions of galaxy
clusters, with similar numbers and properties of dark matter
satellites orbiting within their virial radii.  The amplitude and tilt
of the power spectrum, or varying the cosmological parameters $\Omega$
and $\Lambda$ will have little effect on the abundance of
substructure. These only slightly alter the merger history and
formation timescales. Any difference in merger history will be less
than what we have already explored by comparing the cluster to the
galaxy. Furthermore, we have shown that the properties of the
substructure do not change over a 4 Gyr period, therefore an earlier
formation epoch will not change these results.

If we appeal to gas physics and feedback to hide $95\%$ of the Milky
Way's satellites then we must answer the question why just $5\%$
formed stars with relatively normal stellar populations and reasonably
large baryon fractions. If this problem can be overcome, then the
substructure has several observational signatures, namely disk
heating, gravitational lensing and direct/indirect particle dark
matter detection experiments. Unfortunately, the existence of old thin
disks with no thick/halo components may force us to seek a mechanism
to suppress small scale power {\it e.g.} free streaming by a neutrino
of mass $\sim 1$ keV (Schaeffer \& Silk 1988).



\end{document}